\documentstyle[12pt]{article}
\begin{document}

\hfill DESY 96-155 

\hfill hep-th/9608056 

\hfill August 1996

\vspace*{15mm}

\begin{center}

{\LARGE \bf Complete determination of the singularity structure of 
zeta functions}

\vspace{12mm}

\medskip

 {\sc E. Elizalde}
\footnote{Permanent address: Center for Advanced 
Studies CEAB, CSIC, Cam\'{\i} de Santa B\`arbara,
E-17300 Blanes, and Dept. ECM, Facultat de  F\'{\i}sica,
 Universitat de Barcelona, Diagonal 647,  E-08028
Barcelona, Spain. E-mail: eli@zeta.ecm.ub.es} \\
II. Institut f\"ur Theoretische Physik der Universit\"at Hamburg\\
Luruper Chaussee 149, D-22761 Hamburg, Germany

\vspace{20mm}

{\bf Abstract}

\end{center}

Series of extended
 Epstein type provide  examples of non-trivial zeta functions
with important physical applications. The regular part of their analytic
continuation is seen to be a convergent or an asymptotic series. Their
 singularity structure is completely
  determined in terms of the Wodzicki residue
in its generalized form, which is proven to  yield
 the residua of all the poles of
 the zeta function, and not just that of the rightmost pole
  (obtainable from the Dixmier trace).  
 The  calculation is a very down-to-earth 
application of these powerful functional analytical methods in physics.

\vfill

\noindent {\it PACS:}  11.10.Gh, 02.30.Tb, 02.30.Dk, 02.30.Mv



\newpage


A most important issue in the application of the
 zeta-function regularization method  \cite{zall1} in physics is 
the  precise determination of the pole structre of the 
analytical continuation of the corresponding zeta function. 
In the recent mathematical literature, there are precise 
 results  which partially characterize the meromorphic 
structure of the analytical continuation of the
zeta function of any elliptic pseudodifferential
 operator ($\Psi$DO) \cite{psdo1}, even
of complex order \cite{kv1}.  The position and the order  of the
 poles is known, and also the residue of the rightmost one, which
can be determined by using either the Dixmier 
trace or the Wodzicki
residue of the principal symbol of the operator. 
Here, we  obtain 
the residua of all the remaining poles and 
illustrate their very simple
 calculation through such powerful functional
 analytical tools, by means of two fundamental
 examples with physical application \cite{eli1}. 
The additional determination
that is carried out of the regular part of the analytic
continuation completes the analysis of the 
meromorphic structure of the zeta functions.

A pseudodifferential operator $A$ of order 
$m$ on a manifold $M_n$ is defined through its symbol
$a(x,\xi)$, which is a  function  belonging to the  space 
$S^m(\mbox{\bf R}^n\times \mbox{\bf R}^n)$ 
of $\mbox{\bf C}^\infty$ functions
such that for any $\alpha, \beta$ there 
exists a constant $C_{\alpha,\beta}$ so that
$\left| \partial^\alpha_\xi \partial^\beta_x a(x,\xi) \right| \leq 
 C_{\alpha,\beta} (1+|\xi|)^{m-|\alpha|}$. The definition of $A$
is given (in the distribution sense) by 
\begin{equation}
Af(x) = (2\pi)^{-n} \int e^{i<x,\xi>} a(x,\xi) \hat{f}(\xi) \, d\xi,
\end{equation}
where $f$ is a smooth function ($f \in
 {\cal S}$) and $ \hat{f}$ its Fourier transform.
 When $a(x,\xi)$ is a polynomial in $\xi$
 one gets a differential operator
but, in general,  the order $m$ can be  even  complex.
  Pseudodifferential
operators are useful tools, both in mathematics and in physics.
 They were crucial for the proof of the 
uniqueness of the Cauchy problem \cite{cald57}
and also for the proof of the Atiyah-Singer index formula \cite{as63}. 
In quantum field theory they appear in any
 analytic continuation process (as 
complex powers of differential
operators, like the Laplacian) \cite{bbor}. 
They constitute nowadays the basic starting point
of any rigorous formulation of quantum 
field theory through microlocalization \cite{fre1}, 
a concept that is considered to be
 the most important step towards the understanding 
of linear partial differential equations since the appearance
of distributions.

For $A$ a positive-definite elliptic $\Psi$DO of 
positive order $m \in \mbox{\bf R}$, acting on 
the space of smooth sections of an 
$n$-dimensional vector bundle $E$ over a
 closed, $n$-dimensional
  manifold $M$, the zeta function is defined as
\begin{equation}
\zeta_A (s) = \mbox{tr}\  A^{-s} = \sum_j
 \lambda_j^{-s}, \qquad \mbox{Re}\ s>\frac{n}{m} \equiv s_0.
\end{equation}
Here $s_0$ is called the abscissa of convergence 
of $\zeta_A(s)$, which is proven to have a meromorphic
continuation to  the whole complex plane 
$\mbox{\bf C}$ (regular at $s_0$), 
provided that $A$ admits a spectral cut: $
L_\theta = \left\{  \lambda \in \mbox{\bf C};  
 \mbox{Arg}\, \lambda =\theta, 
\theta_1 < \theta < \theta_2\right\}$,   $\mbox{Spec}\, A
\cap L_\theta = \emptyset$
(the Agmon-Nirenberg condition). Strictly
 speaking, the definition of $\zeta_A (s)$ depends on the
position
of the cut $L_\theta$, not so that of the
 determinant \cite{rs1} $\det_\zeta A = \exp
 [-A'(0)]$, which only depends on the
homotopy class of the cut. The precise structure 
of the analytical continuation is well known \cite{gilk1}: it has
at most simple poles at $s_k = (n-k)/m$, $k=0,1,2,\ldots,n-1,n+1, \dots$. 
The applications of this zeta-function definition 
of a determinant in physics are  important \cite{sz1,wit1}.
 A zeta function with the same
meromorphic structure in the complex $s$-plane 
and extending the ordinary definition 
to operators of complex
order $m\in \mbox{\bf C} \backslash  \mbox{\bf Z}$,
 has been recently obtained in Ref. \cite{kv1}. 
(It is clear that operators of complex order do not admit spectral cuts.)
 The  construction in \cite{kv1} starts from 
 the definition of a trace, obtained as the integral 
over the manifold of the trace density of the difference
between the Schwartz kernel of $A$ and the Fourier 
transform of a number of first homogeneous terms (in
$\xi$) of the usual decomposition of the symbol of
 $A$: $a (x,\xi) =a _m(x,\xi) +a _{m-1}(x,\xi) + \cdots 
+a _{m-N}(x,\xi) + \cdots$

In order to write down an action in operator language 
one needs a functional that replaces integration. For
the Yang-Mills theory this is the Dixmier trace, which
 constitutes
 the unique extension of the usual trace to the ideal
 ${\cal L}^{(1,\infty)} $
of the compact operators  $T $ such that the partial
 sums of its spectrum diverge
logarithmically as the number of terms in the sum, i.e.
\begin{equation}
\sigma_N (T) \equiv \sum_{j=0}^{N-1} \mu_j= {\cal O} 
(\log N), \qquad \mu_0 \geq \mu_1 \geq \cdots
\end{equation}
The definition of the Dixmier trace of $T$, Dtr $T$ 
\cite{dix1}, is then a refinement of the limit 
$\lim_{N\rightarrow \infty}
\frac{1}{\log N} \sigma_N (T)$. (It is directly given 
by this limit when the Cesaro means $M(\rho)$
 of the sequence in $N$ are convergent as $\rho
 \rightarrow \infty$.) As observed by Connes \cite{conn1}, the 
Hardy-Littlewood theorem can be stated in a way
 that connects the Dixmier trace with the residue of the zeta
function of the operator $T^{-1}$ at $s=1$: Dtr $T= 
\lim_{ s \rightarrow  1^+} (s-1) \zeta_{T^{-1}} (s)$. 

The Wodzicki (or noncommutative) residue \cite{wod1}
 is the only extension of the Dixmier trace to 
$\Psi$DOs which are not
in  ${\cal L}^{(1,\infty)} $. Even more, it is the only
 trace at all one can define in the algebra of $\Psi$DOs up to a
multiplicative constant. It is given by the integral 
\begin{equation}
\mbox{res} \ A = \int_{S^*M} \mbox{tr}\   a_n(x,\xi) 
 \, d\xi, \label{wr1}
\end{equation}
with $S^*M \subset T^*M $ the co-sphere bundle on $M$ 
(some authors put a coefficient  in front of the integral). 
If dim $M=n=-$  ord $A$ ($M$ compact Riemann, 
$A$ elliptic, $n\in \mbox{\bf N}$) it coincides with
 the Dixmier trace, and one has \cite{wod1}
\begin{equation}
\mbox{Res}_{s=1} \zeta_A (s) = \frac{1}{n} \, \mbox{res} \ A^{-1}.
\end{equation}
However, the Wodzicki residue continues to make sense 
for $\Psi$DOs of arbitrary order and, even if the symbols
$a_{j} (x, \xi)$, $j<m$, are not invariant under 
coordinate choice, 
the integral in (\ref{wr1}) is, and defines a trace. In
particular, the residua of the poles of the extended 
definition of zeta function 
to operators of complex order are 
also given  by the noncommutative residue.
Moreover, an interesting connection of the 
Wodzicki residue with the second coefficient of the
heat-kernel expansion of the Laplacian has 
been found recently \cite{kw1,ack1}.

A complete determination of the meromorphic
 structure of the zeta  function in the complex plane is
obtained as follows. Relying on the above results, 
what is missing for the description of the
singularities are  the residua of
 all the remaining poles. As for the regular part 
of the analytic
continuation, specific methods have to be used
 and the results are 
 non-trivial: asymptotic series, and not
convergent ones,  appear most often \cite{eli2}. 

\noindent {\it Proposition 1}. Under the conditions 
of existence of the zeta function of $A$, given above,
 and asuming that the
 symbol $a(x,\xi)$ of the operator $A$ is analytic 
in $\xi^{-1}$ at $\xi^{-1}=0$, the
 formula for the determination of the residue of 
the rightmost pole (by means of the
Wodzicki residue) can be generalized  to
  calculate {\it all} the residua of the zeta function
 poles, in the way: 
\begin{equation}
\mbox{Res}_{s=s_k} \zeta_A (s) = \frac{1}{m} \, 
\mbox{res} \ A^{-s_k} =  \frac{1}{m} \int_{S^*M}   \mbox{tr}\ 
a^{-s_k}_{-n} (x,\xi) \, d^{n-1} \xi.
\end{equation}

\noindent {\it Proof}. One just has to notice that the 
homogeneous component of degree $-n$ of the corresponding
power of the principal symbol of $A$ is obtained by taking
 the appropriate derivative of  a power of the symbol with respect
to $\xi^{-1}$ at $\xi^{-1}=0$, namely:
\begin{equation}
a^{-s_k}_{-n} (x,\xi) = \left.  \left( \frac{\partial}{\partial
 \xi^{-1}} \right)^k \left[ \xi^{n-k} a^{(k-n)/m} (x,\xi)
\right] \right|_{ \xi^{-1} =0} \  \xi^{-n}.
\end{equation}
The proof then follows by simple algebraic manipulation.

\noindent {\it Corollary}. As a consequence of the additivity 
property of the trace, it is now clear that, given two operators
$A_1$ and $A_2$ satisfying the conditions above,  the
singularity structure of the zeta function of the operator 
$A_1+A_2$ can be determined from those of the zeta functions
of  $A_1$ and $A_2$. In fact, for all $k$,
\begin{equation}
\mbox{Res}_{s=s_k} \zeta_{A_1+A_2} (s) = 
\mbox{Res}_{s=s_k} \zeta_{A_1} (s) +
 \mbox{Res}_{s=s_k} \zeta_{A_2} (s). 
\end{equation}

These results will now be illustrated with two examples.
Aside from the Riemann and Hurwitz zeta functions,
 the Epstein ones \cite{eps1}
 and  generalizations thereof are 
 most basic tools in the  zeta function regularization
 method  \cite{eli2}.  
They appear in the calculation of the vacuum
 energy or effective potentials of quantum 
physical systems
involving toroidal compactification, finite temperature, 
massive particles, or  a chemical potential. 
Consider a  spacetime with 
topology {\bf R}$\times T^2$ \cite{kke1} and
 a general metric on $T^2$: $ds^2 =h_{ab} dx^a dx^b$,
with
\begin{equation}
h_{ab}=\frac 1 {\tau_2}\left(
   \begin{array}{cc}
     1 & \tau_1\\
      \tau_1 & |\tau|^2
   \end{array}\right) ,
\end{equation}
$(\tau_1, \tau_2)$ being the Teichm\"uller parameters,
  $\tau =\tau_1+i\tau_2$, $\tau_2>0$.
The Laplace-Beltrami operator is:
$L =-\frac 1 {\tau_2} (|\tau|^2 \partial_1^2 -2\tau_1
\partial_1\partial_2
+\partial_2^2 )$
and its eigenvalues
$\lambda_{n_1,n_2} =\frac{4\pi^2}{\tau_2} 
(|\tau|^2n_1^2 -2\tau_1n_1n_2
   +n_2^2 )$.
In the  massive case the spectrum runs over $n_1,n_2 \in$
{\bf Z}. If $m=0$ the zero-mode $n_1=n_2=0$
has to be
excluded. Under different boundary conditions 
(Dirichlet and Neumann, for instance) one gets a restriction
 of the indices to non-negative values and
in many situations ---as is the case of spherical 
compactification---
 a one-dimensional variant of the Laplacian
 appears \cite{eli2}. This leads us to consider  
 two families of  such operators ---plus boundary 
conditions in general. The
corresponding zeta functions belong to the
 family of generalized Epstein zeta functions:
\begin{equation}
\zeta_E(s;a,b,c;q) \equiv
{\sum_{m,n \in \mbox{\bf Z}}}' (am^2+bmn+cn^2+q)^{-s},
\quad \mbox{Re} \, s >1
\label{e1}
\end{equation}
(where $q$ is the mass,  chemical
potential or  finite-temperature contribution) or
 to the more simple version \cite{elif1}:
\begin{equation}
\zeta_G(s;a,c;q) \equiv \sum_{n=-\infty}^{\infty} 
\left[ a(n+c)^2+q
\right]^{-s}, \quad \mbox{Re} \, s >1/2. 
\label{g1}
\end{equation}
The restriction of these  series to non-negative 
values of the indices will be denoted by
 $\zeta_{E_t}$ and $\zeta_{G_t}$,
respectively. 
The parenthesis in  (\ref{e1}) is an
inhomogeneous quadratic form, $Q(x,y)+q$, 
restricted to the  integers. We assume that $a,c >0$ and 
$\Delta =4ac-b^2 >0$
 \cite{cs}.
The starting point for the derivation of the formulas is
Jacobi's  theta function identity 
 \begin{equation}
\sum_{n=-\infty}^{+\infty} e^{-(n+z)^2t} =
 \sqrt{\frac{\pi}{t}} \left[ 1+ \sum_{n=1}^{\infty}
e^{-\pi^2n^2/t } \cos (2\pi n z) \right], \qquad 
  z,t \in \mbox{\bf C}, \ \mbox{Re}\ t>0. \label{tfi4}
\end{equation}

\noindent {\it Example 1}.
Consider first the second function, which is more simple. 
Making use of the Jacobi identity we get the 
 analytical continuation 
\begin{eqnarray}
 \zeta_G(s;a,c;q)  &=&
\sqrt{\frac{\pi}{a}} \, \frac{\Gamma (s-1/2)}{\Gamma (s)} q^{1/2 -s} +
\frac{4\pi^s}{\Gamma (s)} a^{-1/4-s/2} q^{1/4-s/2} \nonumber
\\ && \cdot \sum_{n=1}^\infty
n^{s-1/2} \cos (2\pi nc) K_{s-1/2} (2\pi n\sqrt{q/a}), \label{if1}
\end{eqnarray}
where $K_\nu$ is the modified
Bessel function of the second kind.
Associated with the above zeta functions, but
 considerably more difficult to treat,  are the corresponding
truncated sums, with indices running from 0 to 
$\infty$. In this case the Jacobi identity is of no use. 
By means of specific techniques of analytic continuation
 of zeta functions \cite{eli2}, we obtain 
\begin{eqnarray}
&& \hspace{-4mm} \zeta_{G_t}(s;a,c;q)
 \sim \left(\frac{1}{2} -c \right) q^{-s} + \frac{q^{-s}}{\Gamma (s)}
\sum_{m=1}^{\infty}
\frac{(-1)^m \Gamma (m+s)}{m!} \left( \frac{q}{a} \right)^{-m}
\zeta_H (-2m, c)  \label{if11} \\ &&   +
\sqrt{\frac{\pi}{a}} \, \frac{\Gamma (s-1/2)}{2\Gamma (s)} q^{1/2 -s}
+\frac{2\pi^s}{\Gamma (s)} a^{-1/4-s/2} q^{1/4-s/2}
 \sum_{n=1}^\infty
n^{s-1/2} \cos (2\pi nc) K_{s-1/2} (2\pi n\sqrt{q/a}).  \nonumber
\end{eqnarray}
The first series is
asymptotic \cite{eli11,8}. From the expressions above one can
calculate the determinants
of Klein-Gordon and Dirac operators on compact
 spaces as, for instance,
$N$-cubes, cylinders and spheres $S^N$ (whenever the
spectrum
$\lambda_n$ is  a polynomial in $n$). 

The meromorpic structure of these zeta functions is 
described by the general theory.  According to it,
in principle,  poles at the positions $s=-1,-2,-3, \ldots$ 
could also be possible. They just have
zero residue, as we shall now prove.  The residue of 
 the rightmost pole at $s=1/2$ can be obtained from the
Dixmier trace:
\begin{equation}
\mbox{Dtr}\ G^{-1/2} = \lim_{N\rightarrow \infty}
\frac{1}{\log N} \sum_{j=0}^{N-1}\left[ a(j+c)^2+q
\right]^{-1/2} = \frac{1}{\sqrt{a}},
\end{equation}
which is in fact the value of the residue of the pole 
at $s=1/2$  in Eq. (\ref{if1}), thus Res$_{s=1/2} \zeta_G (s) =$
Dtr $G^{-1/2}$.
Now to the second step: the residue of this pole at 
$s=1/2$ can also be obtained from the Wodzicki
residue. In fact, we have:
\begin{equation}
\mbox{res}\ G^{-1/2} = \int_{S^*R} \mbox{tr}\ 
g^{-1/2}_{-1} (\xi) \, d\xi = \frac{2}{\sqrt{a}}
\end{equation}
(note that $g^{-1/2} (\xi) = \xi^{-1} + {\cal O}
 ( \xi^{-2})$  and that the $0-$dimensional 
sphere is reduced to two
points,  namely $S^*R=S^0 = \{ -1,1\}$). Thus, 
Res$_{s=1/2} \zeta_G (s) = \frac 1 2$
res $G^{-1/2}$.

Having dealt with the rightmost pole in all possible ways, 
we  now analyze  the others by means of the 
generalized Wodzicki residue.  We will
prove that 
\begin{equation}
\mbox{Res}_{s=1/2-k}\ \zeta_G (s) = \frac 1 2 \, 
\mbox{res}\ G^{k-1/2},  \qquad 
\mbox{Res}_{s=-k}\ \zeta_G (s) = 0,  \qquad k=0,1,2, \ldots
\end{equation}
The decomposition of the corresponding symbol
 into homogeneous parts yields
\begin{eqnarray}
 g^{k-1/2} (\xi)& =& \xi^{2k-1} \left[ 1+ \frac{ k-1/2}{ 1!} \frac{ q}{ \xi^2} +
\cdots + \frac{(k-1/2)(k-3/2) \cdots 1/2}{k!} \frac{ q^k}{ 
\xi^{2k}} + \cdots \right] \nonumber \\
& =& \xi^{2k-1}+ \cdots + \frac{ (2k-1)!!}{ k! \, 2^k} \, 
\frac{q^k}{\xi} + \cdots , \qquad k=0,1,2, \ldots,  
\label{deco1} \\
g^{k} (\xi)& =& \left( \xi^{2} +q\right)^k = \xi^{2k} +
 \cdots + q^k, \qquad k=0,1,2, \ldots, \nonumber 
\end{eqnarray}
therefore,
\begin{equation}
 \mbox{tr}\ g^{k-1/2}_{-1} (\xi) = \frac{ (2k-1)!! 
\, q^k }{ k! \, 2^k}, \qquad 
\mbox{tr}\ g^{k}_{-1} (\xi) = 0, \qquad k=0,1,2, \ldots.  \\
\end{equation}
Thus, we obtain 
\begin{equation}
\mbox{Res}_{s=1/2-k}\ \zeta_G (s) =\frac 1 2 \, 
\mbox{res}\ G^{k-1/2}= \frac{ (2k-1)!! \, q^k }{ k! \, 2^k}, \
\mbox{Res}_{s=-k}\ \zeta_G (s) =\frac 1 2 \, 
\mbox{res}\ G^{k}= 0, \ k=0,1,2, \ldots,
\end{equation}
which coincide in fact with the residues of the
 poles of $ \zeta_G (s)$ at $s=s_k$, as we wanted to see.
The fact  that some of the would-be poles are actually not
present follows from the decomposition (\ref{deco1})
  showing  clearly that their residua are zero.
We see in this example how  the Wodzicki residue 
under its more general form allows us to calculate  all the residua
of the poles of the zeta function. The meromorphic 
structure of the analytical continuation of the zeta
 function is
absolutely specified through the Dixmier trace and
 the Wodzicki residue in its general form. However, we
have also  shown through our explicit calculation 
that what remains can  in no way  be considered as a trivial
analytic part. It may  be given by a convergent 
series but, possibly, by an asymptotic one. 

\noindent {\it Example 2}.
It is more involved. In 
 the homogeneous case
the analytic continuation of this Epstein zeta function is
given by the Chowla-Selberg formula \cite{cs}
\begin{eqnarray}
&& \zeta_E(s;a,b,c;0) = 2\zeta (2s)\, a^{-s} + \frac{2^{2s}
\sqrt{\pi}\, a^{s-1}}{\Gamma (s) \Delta^{s-1/2}} \,\Gamma (s
-1/2) \zeta (2s-1)
\nonumber \\ && \hspace{15mm}
+ \frac{2^{s+5/2} \pi^s }{\Gamma (s) \, \Delta^{s/2-1/4}\, \sqrt{a}}
\sum_{n=1}^\infty
 n^{s-1/2}
\sigma_{1-2s} (n) \,
 \cos (\pi n b/a) \,
K_{s-1/2}\left( \frac{\pi n}{a}
\sqrt{ \Delta} \right).
\label{cs1}
\end{eqnarray}
where
$
\sigma_s(n) \equiv \sum_{d|n} d^s,
$
 sum over the $s$-powers of the divisors of $n$. 
(There is a missprint in the transcription of  formula  (\ref{cs1}) in
Ref. \cite{dic}). We observe that the rhs of (\ref{cs1}) exhibits a
simple pole at $s=1$, with residue $
\mbox{Res}_{s=1}  \zeta_E(s;a,b,c;0) = \frac{2\pi}{\sqrt{\Delta}}$.
In the general case, we have obtained the  meromorphic continuation:
 \begin{eqnarray}
&& \hspace{-5mm}  \zeta_E(s;a,b,c;q) = -q^{-s}
+\frac{2\pi q^{1-s}}{(s-1) \sqrt{\Delta}}
 + \frac{4}{\Gamma (s)} \left[
\left( \frac{q}{a} \right)^{1/4}
 \left( \frac{\pi}{\sqrt{qa}} \right)^s
\sum_{k=1}^\infty
k^{s-1/2} K_{s-1/2} \left( 2\pi k \sqrt{\frac{q}{a}} \right) \right.
\nonumber \\ &&  \hspace{6cm} + \sqrt{\frac{q}{a}} \left(2\pi
\sqrt{\frac{a}{q\Delta}} \right)^s
 \sum_{k=1}^\infty k^{s-1} K_{s-1} \left( 4\pi k
\sqrt{\frac{a q}{\Delta}}\right)  \label{cse1}
 \\ && +\left. \sqrt{\frac{2}{a}} (2\pi)^s
 \sum_{k=1}^\infty k^{s-1/2} \cos (\pi k b/a) \sum_{d|k} d^{1-2s} \,
 \left( \Delta + \frac{4aq}{d^2} \right)^{1/4-s/2}
K_{s-1/2}  \left( \frac{\pi k}{a} \sqrt{ \Delta + \frac{4aq}{d^2}}
\right) \right].\nonumber
\end{eqnarray}
This is a fundamental result.  It looks rather different
from the Chowla-Selberg formula (\ref{cs1}), but it can actually be viewed as
its natural extension to the case $q\neq 0$.
It also shares all the good
properties of (\ref{cs1}).

As in example 1, the only pole of this zeta function can be 
obtained by using either the Dixmier trace or the Wodzicki residue. 
In fact, 
\begin{equation}
\mbox{Dtr}\ E^{-1} = \lim_{N\rightarrow \infty}
\frac{1}{\log  (N^2)} \sum_{m,n =-N}^N (am^2+bmn+cn^2+q)^{-1} 
 = \frac{2 \pi}{\sqrt{\Delta}},
\end{equation}
which is  the value of the residue of the pole at $s=1$
, thus Res$_{s=1} \zeta_E (s) =$
Dtr $E^{-1}$. Moreover, this value can also be obtained 
from the Wodzicki
residue:
\begin{equation}
\mbox{res}\ E^{-1} = \int_{S^*\mbox{\bf R}^2} 
\mbox{tr}\ e^{-1}_{-2} (\xi) \, d^2\xi = 
\frac 1 a \int_0^{2\pi} \left[ \left( \tan \theta +
\frac{b}{2a} \right)^2 
+ \frac{\Delta}{4a^2} \right]^{-1}  d(\tan\theta) 
= \frac{4\pi}{\sqrt{\Delta}}.
\end{equation}
Here the integral is performed over the unit
 circumference ($S^*\mbox{\bf R}^2 =S^1, |\xi| =1$).
 Thus, Res$_{s=1} \zeta_E (s) = \frac 1 2$ res $E^{-1}$.
 We have thus shown again how the rightmost
pole of the zeta function can be obtained either from 
the Dixmier trace or from the Wodzicki residue.  The fact
that this is the  only pole of our zeta function  also follows
 from the calculation of the generalized Wodzicki
 residua. According to the general theory, the other 
possible poles would be at $s=s_k=1-k/2$, $k=1,3,4,\ldots$. 
We must obtain the homogeneous component of
 degree $-2$ of the principal symbol of the operator
$E^{k/2-1}$:
\begin{equation}
e^{k/2-1} (\xi_1,\xi_2) = \left( a\xi_1^2 +b \xi_1\xi_2
 + c \xi_2^2 +q \right)^{k/2-1}.
\end{equation}
But it is  clear that neither for $k$ odd nor for $k$
 even is there any component of this  principal symbol of
degree $-2$. All corresponding residua are zero
 and none of these poles exists.

The most difficult case in the family of Epstein-like
 zeta functions corresponds to having 
a truncated range. This
comes about when one imposes boundary conditions 
of the usual Dirichlet or Neumann type \cite{eli2}.
 Jacobi's theta function identity is then useless
 and no expression in terms of a 
convergent series for the analytical continuation to
 values of Re $s$ below the abscissa of convergence 
can be obtained. The best one gets
 is an asymptotic series expression. However, the issue of extending
 the Chowla-Selberg formula or,  better still, the 
more general one we have obtained before for
 inhomogeneous
 Epstein zeta functions in two indices, is not 
simple and had never 
been adressed in the literature. 
In order to obtain the analytic continuation to  
  Re $s  \leq 1$  of the
 truncated inhomogeneous Epstein zeta 
function in two dimensions,
$ \zeta_{E_t}(s;a,b,c;q) \equiv \sum_{m,n =0}^\infty 
(am^2+bmn+cn^2+q)^{-s} $,
we can  proceed in two ways: either by a direct 
calculation that leads to the generalized 
Chowla-Selberg formula \cite{eli2}
or by using the formulas for the  Epstein zeta 
function in  one dimension (example 1)
 recurrently. In both cases the result  is
 \begin{eqnarray}
&& \hspace{-5mm}   \zeta_{E_t}(s;a,b,c;q) 
 \equiv \sum_{m,n =0}^\infty 
(am^2+bmn+cn^2+q)^{-s}   \nonumber \\
&&  \hspace{-2mm}  \sim \, \frac{(4a)^s}{\Gamma (s)} \sum_{m,n =1}^\infty 
\frac{(-1)^m \Gamma (m+s)}{m!} (2a)^{2m} (\Delta n^2 +4aq)^{-m-s}
\zeta_H \left(-2m; \frac{bn}{2a} \right)  \nonumber \\
&& -\frac{b\, q^{1-s}}{(s-1)\Delta \Gamma (s-1)}
 \sum_{n =0}^\infty \frac{(-1)^n \Gamma (n+s-1)B_n}{n!}
 \left( \frac{4aq}{\Delta} \right)^{-n} \nonumber \\
&&  + \frac{q^{-s}}{4}
 +\frac{\pi q^{1-s}}{2(s-1) \sqrt{\Delta}}  + \frac{1}{4} \left( 
\sqrt{\frac{\pi}{a}} + \sqrt{\frac{\pi}{c}} \right)
\frac{\Gamma (s-1/2)}{\Gamma (s)} q^{1/2-s}  \nonumber \\
&&  
 + \frac{1}{\Gamma (s)} \left[ 2
\left( \frac{q}{a} \right)^{1/4}
 \left( \frac{\pi}{\sqrt{qa}} \right)^s
\sum_{k=1}^\infty
k^{s-1/2} K_{s-1/2} \left( 2\pi k \sqrt{\frac{q}{a}} \right) \right.  \label{cser1}
 \\ &&\hspace{-8mm}  +\left( \frac{aq}{\Delta} \right)^{1/4}
 \left( \pi \sqrt{\frac{a}{q\Delta}} \right)^s
\sum_{k=1}^\infty
k^{s-1/2} K_{s-1/2} \left( 2\pi k \sqrt{\frac{aq}{\Delta}} \right) 
+ \sqrt{\frac{q}{a}} \left(2\pi
\sqrt{\frac{a}{q\Delta}} \right)^s
 \sum_{k=1}^\infty k^{s-1} K_{s-1} \left( 4\pi k
\sqrt{\frac{a q}{\Delta}}\right)  \nonumber
 \\ && +\left. \sqrt{\frac{2}{a}} (2\pi)^s
 \sum_{k=1}^\infty k^{s-1/2} \cos (\pi k b/a) \sum_{d|k} d^{1-2s} \,
 \left( \Delta + \frac{4aq}{d^2} \right)^{1/4-s/2}
K_{s-1/2}  \left( \frac{\pi k}{a} \sqrt{ \Delta + \frac{4aq}{d^2}}
\right) \right].\nonumber
\end{eqnarray}
This imposing formula is new too.
 The first series is in general 
asymptotic, but it converges for a  wide range of values of
the parameters. The second series is always asymptotic and
contributes to the pole at $s=1$. As in the case of
 the fundamental formula,
Eq. (\ref{cse1}), the pole structure is here
explicitly, although  much more elaborate. 
Apart from the pole at $s=1$,  there is here a sequence of poles
at $s=\pm 1/2, -3/2, -5/2, \ldots$.
Calculations  similar to the ones above lead to
  the determination of the residua of the poles in 
this case by means of the same
expressions as before.  

\vspace{3mm}

\noindent{\bf Acknowledgments}.
The author is indebted with Klaus
Fredenhagen and Romeo Brunetti for enlightening
discussions and with all the members of the II. Institut
 f\"ur Theoretische Physik der Universit\"at Hamburg  for
the very kind hospitality. 
 This investigation has been  supported by 
 the  Alexander-von-Humboldt Foundation, 
DFG-CSIC,  DGICYT and CIRIT.


\end{document}